\documentclass[aip,jmp,amsmath,amssymb,
 reprint,%
]{revtex4-1}

\usepackage{graphicx}
\usepackage{dcolumn}
\usepackage{bm}
\usepackage[mathscr]{eucal}
\usepackage{amsopn}
\usepackage{amsthm}
\usepackage{color}

\theoremstyle{definition}
\newtheorem{Remark}{Remark}

\DeclareMathOperator{\CP}{\mathbb{C}P}
\DeclareMathOperator{\Real}{\mathbb{R}}
\DeclareMathOperator{\Complex}{\mathbb{C}}
\DeclareMathOperator{\Ibb}{\mathbb{I}}
\DeclareMathOperator{\Hilbert}{\mathcal{H}}
\DeclareMathOperator{\diag}{diag}
\DeclareMathOperator{\Tr}{Tr}

\definecolor{darkgreen}{rgb}{0,0.4,0}

\begin{document}

\preprint{}

\title[Reconstruction of Hamiltonians from  given time evolutions]{Reconstruction of Hamiltonians from  given time evolutions}

\author{J. Bernatska}
\email{BernatskaJM@ukma.kiev.ua}
\affiliation{
National University of `Kiev-Mohyla Academy', 2 Skovorody Str., 04070 Kiev, Ukraine}
\altaffiliation[Also at ]{Bogolyubov Institute for Theoretical Physics.}

\author{A. Messina}
\email{messina@fisica.unipa.it}
\affiliation{Dipartimento di Fisica, Universit\`{a} di Palermo, Via Archirafi 36,
90123 Palermo, Italy}

\date{\today}

\begin{abstract}
In this paper we propose a systematic method to solve the inverse
dynamical problem for a finite-level quantum system governed by the von Neumann
equation: to find a class of Hamiltonians reproducing a
prescribed time evolution of a pure or mixed state of the system.
Our approach exploits the equivalence between the action of the
group of evolution operators over the state space and the adjoint
action of the unitary group over Hermitian matrices. The method is
illustrated by two examples involving a pure and a mixed state.
\end{abstract}

\pacs{02.20.Qs, 03.65.Aa}

\keywords{pure and mixed states, von Neuman equation, unitary group, adjoint action, stereographic parametrization} 
\maketitle

\section{Introduction}
The problem of reconstructing a Hamiltonian or a Lagrangian from
observations on a physical system is a fundamental issue in physics.
Its history can be traced back to the time of Kepler and Newton when
the law of gravitation was discovered from observations of the
planetary motion. The reconstruction of an effective potential from
the data of particle scattering also belongs to this kind of problems.
In the context of quantum control theory
reconstructing Hamiltonians on demand represents a strategic intermediate step
toward the realization of noise-protected control
protocols governing the dynamics of a given, often
complex and multipartite, quantum system.

To drive the time evolution of a quantum system at will is at the
heart of the quantum control theory. Fruitful applications of this
theory in physical chemistry allow to achieve the optical control
over ultrafast molecular processes \cite{bonacic}, and to design
feedback control for quantum state preparations in quantum optics
\cite{handel}. A specific task of the quantum control theory is to
contrive schemes that allow to steer a quantum system to a desired
state starting from a given initial one. More ambitiously one could
even require a possibility to control an evolution of the system
under scrutiny at any instant of time. In other words, one could
require to design the control in such a way that the system evolves
along a prefixed path in its state space. Being
able to solve this problem, a dream since the early days of quantum
mechanics, would help to highlight dynamical
behaviors of microscopic systems, and would  provide additional tools and ideas applicable to quantum
computation. In a recently published paper \cite{Leggio} the
tomographic method has been used to show that a specially modeled
time-dependent state of two qutrits, violating Bell's inequalities
at $t\,{=}\,0$, loses this non-classical feature at successive
times. In this paper we introduce a systematic
method to solve the following inverse problem of the unitary quantum
dynamics: given the time evolution of a system is explicitly specified
by means of a density matrix $\rho$, to construct a class of generally time-dependent Hamiltonian
models whose associated von Neumann equations are satisfied by
this~$\rho$, describing a pure or a mixed quantum
state.

The paper is organized as follows. In Section 2 mathematical
notations are introduced. In Section 3 we explain our mathematical
procedure: how to reconstruct the
Hamiltonian reproducing the prescribed
time evolution, if this evolution obeys the von Neumann
equation. Every time evolution is realized by a class of Hamiltonian models.
We illustrate our method in
Section 4 where it is applied to time evolutions of both a pure
state and a mixed one.

\section{Mathematical notation}
To establish our notation we start with brief recalling the mathematical
grounds of quantum mechanics after J. von Neumann\cite{Neumann,Emch}.
A quantum system is generally described by three basic ingredients: states, observables,
and dynamics (the law of time evolution). Very often the state space is identified with a complex Hilbert space~$\Hilbert$.
In what follows we consider finite-level systems, so the Hilbert space is finite dimensional. After Dirac we denote its vectors  by $|\Phi\rangle$, $|\Psi\rangle$, \ldots, and an inner product by~$\langle \Phi | \Psi \rangle$.

But the Hilbert space  represents only pure states.
Every vector $|\Phi \rangle\,{\in}\,\Hilbert$ gives rise to the ray (the one-dimensional subspace) $\{\lambda |\Phi \rangle \,{\mid}\, \lambda\,{\in}\,\Complex\}$, associated with one pure state, say $\varphi$. In other words, a pure state is identified with an equivalence class of vectors of length~1 in~$\Hilbert$, and two vectors represent the same pure state if they differ only by a phase factor.

If we want to include mixed states in our pattern, it is better to represent  the pure state
$\varphi$ as a  projection operator $P_{\Phi}$ onto
the ray $\{\lambda |\Phi \rangle \,{\mid}\,\lambda\,{\in}\,\Complex\}$.  In~Dirac's notation
the operator $P_{\Phi}$ is expressed as $|\Phi
\rangle \langle \Phi|$. Every such  projector $P_{\Phi}$, and so every pure state $\varphi$,  is a self-adjoint linear transformation on the Hilbert space such that  $P_{\Phi}^2\,{=}\,P_{\Phi}$. The~collection of all pure states form a projective space.
A `statistical mix' of pure states \begin{equation*}
\rho = \sum_k p_k P_{\Phi_k},
\end{equation*}
where $0\,{\leqslant}\, p_k \,{<}\, 1$ and $\sum_k p_k
\,{=}\, 1$, represents a mixed state. The operator $\rho$  is not a projector, but a self-adjoint positive definite linear transformation. Therefore the state space of a quantum system can be identified with the subspace of positive definite matrices  in the space $\mathfrak{A}(\Hilbert)$ of self-adjoint linear transformations on~$\Hilbert$. In order to identify a quantum state with its density matrix we restrict the state space to matrices  of trace 1.

We introduce a scalar product in $\mathfrak{A}(\Hilbert)$, and the dual space to $\mathfrak{A}(\Hilbert)$ as follows:
\begin{equation*}
(\rho, A) \equiv \Tr \rho A,\qquad \rho \in \mathfrak{A}(\mathcal{H}),\  \  A\in \mathfrak{A}^{\ast}(\mathcal{H}).
\end{equation*}
The dual space $\mathfrak{A}^{\ast}(\mathcal{H})$ represents the associative  algebra of observables\footnote{Associating the space $\mathfrak{A}(\Hilbert)$ with quantum states and the dual space $\mathfrak{A}^{\ast}(\mathcal{H})$ with observables we are guided by their laws of time evolution. Namely, the evolution of a state $\rho$ is realized by the adjoint action $U_{t,t_0} \rho(t_0) U_{t,t_0}^{-1}$, and the evolution of an observable $A$ is realized by the coadjoint action $U_{t,t_0}^{-1} A(t_0) U_{t,t_0}$.}
of a quantum system.  The expectation value of an observable $A$
with respect to a state $\rho$ is defined by the formula
\begin{equation*}
\langle A \rangle_{\rho} = (\rho,A) \equiv \Tr \rho A.
\end{equation*}
Then for a pure state $\varphi\,{=}\,P_{\Phi}$ we have
\begin{equation*}
\langle A \rangle_{\varphi} = \Tr P_{\Phi} A =  \langle \Phi  \,{\mid}\, A \Phi \rangle.
\end{equation*}

The last ingredient of a quantum system is its dynamics, which is governed
by the well-known Schr\"{o}dinger equation
\begin{equation*}
i\hbar \frac{d |\Phi \rangle }{dt}= H |\Phi \rangle,\qquad  |\Phi \rangle \in \Hilbert,
\end{equation*}
if we deal with the pure state associated with the vector~$|\Phi\rangle$; or in general by
 the von Neumann equation
\begin{equation*}
i\hbar \frac{d \rho}{dt}  = [H,\rho],\qquad \rho \in  \mathfrak{A}(\mathcal{H}),
\end{equation*}
where $[\cdot,\cdot]$ denotes a commutator. The self-adjoint operator $H\,{\in}\,\mathfrak{A}^{\ast}(\mathcal{H})$
is called a \emph{Hamiltonian}.

According to Stone's theorem the time evolution generated by a time-independent self-adjoint operator $H$ is realized by
unitary operators $\{U_t\}$, forming a continuous one-parameter group in $t$, that is $U_{t}U_{s}\,{=}\,U_{t+s}$ for all $t,\,s\,{\in}\, \Real$ and $\lim_{t\to 0} U_t \,{\to}\, \Ibb$. Then the  Schr\"{o}dinger equation with an initial condition at $t_0$ is solved by the formula $|\Phi (t)\rangle \,{=}\, U_{t-t_0} |\Phi(t_0)\rangle$.  In the case of  a time-dependent $H$ time evolution is also realized by a continuous group of unitary operators\cite{Messiah,Woodhouse} $\{U_{t,t_0}\}$ such that $U_{t_2,t_1} U_{t_1,t_0}\,{=}\,U_{t_2,t_0}$, $U_{t_0,t_0}\,{=}\,\Ibb$; and the  Schr\"{o}dinger equation is solved by  $|\Phi (t)\rangle \,{=}\, U_{t,t_0} |\Phi(t_0)\rangle$.

Hereinafter we deal with a more general pattern, where the state space is identified with  the subspace $\mathcal{P}$ of positive definite matrices of trace 1 in the space $\mathfrak{A}(\Hilbert)$, and dynamics is governed by the von Neumann equation. Time evolution is realized by the group of unitary operators $\{U_{t,t_0}\}$ as follows
\begin{equation}\label{EvolDM}
\rho(t) = U_{t,t_0} \rho(t_0) U_{t,t_0}^{-1}.
\end{equation}
The Hamiltonian generating this evolution relates to the evolution operator $ U_{t,t_0}$
according to the formula
\begin{equation}\label{HamDef}
H(t)\,{=}\, i\hbar \frac{d U_{t,t_0}}{dt}\, U_{t,t_0}^{-1},
\end{equation}
which can be obtained by substituting the evolution of
$|\Phi\rangle$, or $\rho$, into the corresponding differential equation.

The mentioned well-known facts of quantum mechanics are naturally interpreted
from the position of the Lie group and algebra theory.
Evidently, time evolution of a quantum system is realized by the adjoint action \eqref{EvolDM}
of a unitary group. This gives us a hint how to treat the problem.
Suppose we deal with a $n$-level system, then the state space
can be identified with the positive definite subspace $\mathcal{P}$ of the affine linear space $i\mathfrak{su}(n)\,{\oplus}\,(1/n)\hat{\Ibb}_n$.  In our context the latter space can be replaced by the algebra $i\mathfrak{su}(n)$, what
 we show in the next section.
Therefore, from the mathematical point of view we deal with the adjoint action of a Lie group over its Lie algebra.

Moreover, it is easy to check that $\Tr
\rho^k$ remains invariant under time evolution for any
$k\,{\geqslant}\, 1$:
\begin{equation}\label{OrbitEqs}
\Tr \rho^k(t)\,{=}\,\Tr \rho^k\,(0) = \text{const},
\end{equation}
where only first $n\,{-}\,1$ equations are functionally independent for a $n$-level system.
In other words,
the adjoint action \eqref{EvolDM} divides the state space into non-intersecting orbits, and
every orbit is determined by the set of equations \eqref{OrbitEqs}.
This clarifies the geometry of  time evolution of a closed quantum system.
Every quantum state evolves within a single orbit, fixed by an initial state. It means:
only states within this orbit are reachable from the initial state.

\section{Scheme of reconstruction for $n$-level systems}
We aim at solving the problem of reconstructing a Hamiltonian
model from a given time evolution
satisfying the von Neumann equation.
Actually we reconstruct an evolution operator, which is
considered as an element\footnote{More precise, the proposed method of stereographic parametrization allows to
reconstruct an element of a coset space representing an adjoint orbit.} of a unitary group. For this
purpose we use  a generalized stereographic parametrization\cite{Skrypnyk, BernatskaGIQ}, applicable to
states with finite number of components, say $n$.  Therefore in
what follows we deal with matrices of size $n\,{\times}\, n$, and
mark them by the hat sign $\hat{\ }$.

\subsection{An orbit structure of the state space}\label{ss:OrbitStruct}
First of all, we show an equivalence between the described above physical and mathematical patterns.
Let us consider the algebra $i\mathfrak{su}(n)$ consisting of Hermitian traceless matrices. We denote
these matrices by $\hat{\mu}$, keeping $\hat{\rho}$ for matrices of trace~1. The
adjoint action of the unitary group $\mathrm{SU}(n)$ over  $i\mathfrak{su}(n)$ is naturally defined by
\begin{equation}\label{AdjActionUG}
\hat{\mu}\,{=}\,\hat{u} \hat{\mu}_{\rm in} \hat{u}^{-1}, \qquad
\hat{\mu}\,{\in}\,i\mathfrak{su}(n), \quad \hat{u}\,{\in}\,\mathrm{SU}(n).
\end{equation}
All $\mu$ obtained from the initial point $\hat{\mu}_{\rm in}$ when $\hat{u}$ runs the group  $\mathrm{SU}(n)$
form an orbit.  This orbit can be determined by the set of equations
\begin{equation}\label{OrbitEqsAlg}
\Tr \hat{\mu}^k\,{=}\, \Tr \hat{\mu}_{\text{in}}^k = \text{const},\quad  k=1,\,\dots\, n-1.
\end{equation}
The whole algebra $i\mathfrak{su}(n)$ is devided  into non-intersecting orbits by the adjoint action of $\mathrm{SU}(n)$.

It is easy to check that the adjoint action over $i\mathfrak{su}(n)$ in \eqref{AdjActionUG} and the time evolution over the state space~$\mathcal{P}$ in  \eqref{EvolDM} are realized by the same unitary group. It means that for a $n$-level quantum system the group of unitary operators $\{U_{t,t_0}\}$ is $\mathrm{SU}(n)$.
Also the algebra $i\mathfrak{su}(n)$ has the same orbit structure as the affine space $i\mathfrak{su}(n)\,{\oplus}\, (1/n)\hat{\Ibb}_n$ containig the state space of the system. Indeed, every orbit in $i\mathfrak{su}(n)$ determined by \eqref{OrbitEqsAlg} corresponds to a unique orbit \eqref{OrbitEqs} in $i\mathfrak{su}(n)\,{\oplus}\, (1/n)\hat{\Ibb}_n$. Therefore instead of the affine space we consider the Lie algebra $i\mathfrak{su}(n)$. Also we shift the state space~$\mathcal{P}$ into the corresponding subspace $\mathcal{M}\,{\subset}\,i\mathfrak{su}(n)$ by subtracting the matrix $(1/n)\hat{\Ibb}_n$.

Now we can use the knowledge of the orbit structure of a Lie algebra under the adjoint action of its Lie group in order to
describe the structure of the state space. Above we called this structure the geometry of time evolution.
Every orbit in $i\mathfrak{su}(n)$ is specified by its initial point taken from a fundamental Weyl chamber or its walls. The Weyl chamber is a subset of the Cartan subalgebra $\mathfrak{h}$ of $i\mathfrak{su}(n)$, which is defined as
the maximal Abelian subalgebra of a semisimple Lie algebra. For $i\mathfrak{su}(n)$ it consists of real diagonal matrices of zero trace. Thus a canonical form of $\hat{\mu}\,{\in}\,i\mathfrak{su}(n)$ can serve as an initial point for the orbit where $\hat{\mu}$ is located. But there is an ambiguity upon the order of diagonal entries. This  ambiguity is removed by introducing the Weyl group, which performs a permutation of the diagonal entries, and whose fundamental domain is called a fundamental Weyl chamber.

Here we choose the positive Weyl chamber as a fundamental one. Its interior is the open domain $$C = \{\hat{\mu} \in \mathfrak{h} \mid \langle \hat{\mu},\alpha \rangle >0, \, \forall \alpha \in \Delta^{+}\},$$ where $\Delta^{+}$ denotes the set of positive simple roots.
A wall of the positive Weyl chamber is the set  $$\Gamma_{\alpha} \,{=}\, \{\hat{\mu}\,{\in}\,
\mathfrak{h} \mid \langle \hat{\mu},\alpha \rangle = 0\}, \quad \alpha \in \Delta^{+}$$  or any intersection of such sets.

We omit the accurate definition of simple roots, instead we represent them in the standard for the algebra $i\mathfrak{su}(n)$ form. The simple roots belong to the dual subalgebra $\mathfrak{h}^{\ast}$ and form a basis in it. Because $\mathfrak{h}^{\ast}$ is equivalent to $\mathfrak{h}$ we can use the same representation. The set~$\Delta^{+}$ consists of $n\,{-}\,1$ diagonal matrices $\{\hat{\alpha}_{k}\,{=}\,\hat{E}_{kk}-\hat{E}_{k+1,k+1}\}$, where the matrix $\hat{E}_{jk}$ has the only nonzero entry in row~$j$ column~$k$ and this entry equals~1.
 It is easy to construct the dual basis $\{\hat{\gamma}^{(k)}\}\,{\subset}\,\mathfrak{h}$:
\begin{equation*}
\hat{\gamma}^{(k)} = \sum_{j=1}^k \hat{E}_{jj} - (k/n)\hat{\Ibb}_n.
\end{equation*}
Then every point of the positive Weyl chamber of  $i\mathfrak{su}(n)$ can be represented in the form
\begin{equation*}
\hat{\mu}_{{\rm in}}  = \sum_{k=1}^{n-1} b_k  \hat{\gamma}^{(k)}, \quad b_k \in \Real^{+}.
\end{equation*}
If we allow the coefficients $\{b_k\}$ to equal zero, we obtain the closed  positive Weyl chamber containig all its walls. Every wall spans a nontrivial subset of $\{\hat{\gamma}^{(k)}\}$, and  is represented by a hyperplane bounding the chamber.

Points from the interior of a fundamental Weyl chamber in $i\mathfrak{su}(n)$
give rise to generic orbits of dimension $n^2\,{-}\,n$. Every point of a wall of the Weyl chamber gives rise to a
degenerate orbit of lower dimension. In the case of a one-dimensional wall one obtains a maximal
degenerate orbit of dimension $2(n\,{-}\,1)$. The latter represents a set of pure states if being shifted back into the space $i\mathfrak{su}(n)\,{\oplus}\, (1/n)\hat{\Ibb}_n$ remains a projective space.

In particular,  the points  $b_1\hat{\gamma}^{(1)}$ and $b_{n-1}\hat{\gamma}^{(n-1)}$ give rise to maximal degenerate orbits in $\mathcal{M}$ if $b_1\,{\leqslant}\,1$, $b_{n-1}\,{\leqslant}\,1/(n\,{-}\,1)$. But only one of them, namely $\hat{\gamma}^{(1)}$, generates the set of pure states. Positive definiteness of the state space $\mathcal{P}$ implies that acceptable values of $\{b_k\}$ satisfy the inequalities
\begin{equation*}
n\sum_{k=j}^{n-1} b_k  -  \sum_{k=1}^{n-1} k b_k + 1  \geqslant 0,\quad  j=1,\dots, n.
\end{equation*}
The last inequality ($j\,{=}\,n$) appears to be sufficient  if one takes into account non-negativity of $\{b_k\}$.

Obviously, the elements of the basis $\{\hat{\gamma}^{(k)}\}$, except $\hat{\gamma}^{(1)}$, do not belong to the subspace~$\mathcal{M}$. We are able to construct a basis  in $\mathfrak{h}$ fulfiling this condition. This new basis consists of  traceless real diagonal matrices~$\{\hat{\pi}^{(k)}\}$ such that $(\hat{\pi}^{(k)})^2 \,{-}\, \hat{\pi}^{(k)}\,{=}\,0$.   It is easy to write these matrices explicitely:
\begin{equation*}
\hat{\pi}^{(k)} = \hat{E}_{kk} - (1/n)\hat{\Ibb}_n.
\end{equation*}
The basis $\{\hat{\pi}_k\}$ consists of projectors,
and the expansion
\begin{equation}\label{PrForm}
\hat{\mu}_{{\rm in}} =  \sum_{k=1}^{n-1} c_k \hat{\pi}^{(k)}
\end{equation}
 immediately displays a dimension of the orbit passing through $\hat{\mu}_{{\rm in}}$. The dimension is $(n\,{-}\,1)(m\,{+}\,1)$,
where $m$ denotes a number of nonzero summands.
Simply speaking the degeneracy of an orbit becomes evident from coincidence of diagonal entries of~$\hat{\mu}_{\text{in}}$. For example, if all but one entries are equal, then the orbit passing through $\hat{\mu}_{\text{in}}$ is maximal denerate.

Therefore, every evolution of a closed $n$-level quantum system is located within a certain orbit of the adjoint action of the group $\mathrm{SU}(n)$. The orbit is fixed by an initial state, say $\hat{\rho}(t_0)$. It means that only the states within the fixed orbit are reachable from~$\hat{\rho}(t_0)$ because orbits do not intersect.  All orbits together form the state space~$\mathcal{P}$.

Turning back to our problem, suppose we are given a sample of evolution $\hat{\rho}$ as a function in $t$, and we aim at constructing an operator $\hat{U}_{t,t_0}\,{\in}\,\mathrm{SU}(n)$  realizing the given evolution. First we should identify the orbit where $\hat{\rho}$ lives. We propose to identify it by the canonical form of $\hat{\rho}$, say $\hat{\rho}_{\rm in}$. The matrix $\hat{\rho}_{\rm in}$ is constant and characterizes the whole orbit.  In particular, it serves as a  canonical form for $\hat{\rho}(t_0)$. The constancy of $\hat{\rho}_{\text{in}}$ follows from the set of equations \eqref{OrbitEqs} determining an orbit. Indeed, fixed values of $\Tr \hat{\rho}^k$ guarentee that all poins of an orbit have the same characteristic polynomial, and so have the same eigenvalues. In order to detect a  degeneracy of the orbit, one should apply the expansion \eqref{PrForm} to the matrix $\hat{\rho}_{\rm in}\,{-}\,(1/n)\hat{\Ibb}_n$, that displays a dimension of the space where the given state $\hat{\rho}$ evolves. This determines a structure of the evolution operator $\hat{U}_{t,t_0}$ and some auxiliary matrices in the proposed method.

The next step of our scheme consists in constructing a generic form
of the unitary matrix $\hat{u}\,{\in}\,\mathrm{SU}(n)$
realizing motion within a chosen orbit. We call this matrix an `evolution' one,  indicating by quotes the difference between a motion and a time evolution realised by an evolution operator $\hat{U}_{t,t_0}$.

\subsection{Stereographic parametrization and `evolution' operators}\label{ss:StereogrProj}
As shown above, motion of a $n$-level quantum system  is realized within an orbit by elements of the unitary
group~$\mathrm{SU}(n)$. We call these elements
`evolution' matrices when they concerns motion within orbits, and evolution operators when they generate time evolution.
There is a connection between  an orbit and its `evolution' matrix.  An `evolution' matrix can be written
in terms of the complex coordinates parameterizing an orbit. In the paper we implement this  by the method of generalized stereographic parametrization.

First we briefly recall the idea of the generalized stereographic
parametrization\cite{Skrypnyk,BernatskaGIQ} for a compact Lie group $G$.
It consists in comparing  the
Iwasawa and Gauss decompositions of the complexified group
$G^{\Complex}$ whose maximal compact subgroup is $G$. Let $T$ be the
maximal torus of $G$, and $T^{\Complex}$ be the maximal torus of
$G^{\Complex}$. We also define the following subgroups of
$G^{\Complex}$: the maximal real Abelian subgroup $A$, a nilpotent
subgroup~$R$, and another nilpotent subgroup $Z\,{=}\,R^{\ast}$.
Note that $T^{\Complex}\,{=}\,TA$. The Iwasawa and Gauss
decompositions of $G^{\Complex}$, respectively, are
\begin{equation*}
G^{\Complex}=GAR,\qquad G^{\Complex}= Z T^{\Complex}R.
\end{equation*}
If we define a generic adjoint orbit by means of coset space we obtain the equality line
\begin{equation*}
\mathcal{O} \equiv \frac{G}{T} = \frac{G^{\Complex}}{P} \sim \frac{ZT^{\Complex}R}{T^{\Complex}R} \sim Z,
\end{equation*}
where $P\,{=}\,TAR$ is the minimal parabolic subgroup of
$G^{\Complex}$. Therefore, we can parameterize the orbit
$\mathcal{O}$ in terms of canonical coordinates of $Z$, we denote them $\{z_k\}$.

In general an orbit of the adjoint action of group $G$ is defined as the coset space $G/G_{\rho_{\rm in}}$, where  $G_{\rho_{\rm in}}$ denotes the stability subgroup of the point $\hat{\rho}_{\text{in}}$ identifying the orbit.
A~generic orbit arises when a stability subgroup coincides with the maximal torus $T$.
A larger stability subgroup gives rise to a degenerate orbit, then a nonminimal parabolic subgroup $P\,{=}\,G_{\rho_{\rm in}}AR$ should be taken. In this case the nilpotent subgroup $Z$ is generated by a smaller number of root vectors, and is parameterized by a smaller number of canonical coordinates.

Here we deal with the group $\mathrm{SU}(n)$. Its complexification
having $\mathrm{SU}(n)$ as a maximal compact subgroup is
$\mathrm{SL}(n,\Complex)$. All mentioned subgroups can be
represented in matrix form. The Abelian subgroup $A$ consists of
diagonal matrices $\hat{a}\,{=}\,\diag(d_1,\,d_2,\,\dots,\, d_n)$
with $\prod_{k} d_k\,{=}\,1$. Let $R$ be a subgroup of upper
triangular matrices $\hat{r}$, then $Z$ is a subgroup of lower
triangular matrices $\hat{z}$.

For a matrix $\hat{z}\,{\in}\,Z$, assigned to a point of an
orbit, we obtain an `evolution' matrix $\hat{u}\,{\in}\,
\mathrm{SU}(n)$ from the Iwasawa decomposition
of $\hat{z}$:
\begin{equation}\label{IwasawaDec}
\hat{z}\,{=}\,\hat{u}\hat{a}\hat{r}.
\end{equation}
At the same time we aim at expressing
the matrix $\hat{u}$  in terms of the canonical coordinates $\{z_k\}$ apperaing as entries of the matrix $\hat{z}$.
The easiest way to achieve this goal is to solve the equation
\begin{equation}\label{IwasawaCalc} \hat{z}^{\ast} \hat{z} =
\hat{r}^{\ast} \hat{a}^{\ast} \hat{u}^{\ast} \hat{u} \hat{a} \hat{r}
= \hat{r}^{\ast} \hat{a}^2 \hat{r}.
\end{equation}
for all entries of $\hat{a}$ and $\hat{r}$.
Having the matrices $\hat{a}$ and $\hat{r}$ expressed by means of $\{z_k\}$,
one easily calculates the matrix $\hat{u}$ from the equality \eqref{IwasawaDec}.

As a result, we obtain a generic form for the matrix $\hat{u}$,
parameterized by the complex canonical  coordinates~$\{z_k\}$. This matrix
realizes the `evolution' starting from zero-point $\{z_k\,{=}\,0\}$  within every orbit of the same type.

\begin{Remark}\label{r:1}
It should be noticed, that the obtained `evolution' matrix $\hat{u}$
is not a unique solution of the problem. One can multiply it by any matrix $\hat{v}\in G_{\rho_{\rm in}}$,
that gives another `evolution' within the same orbit.
\end{Remark}

\subsection{Pure states}
Here we consider how to apply the above approach to the
particular case of pure states. Usually such states are described
by vectors of the Hilbert space $\Hilbert$. Let $\{|k\rangle\}$
denote a set of basis quantum states, then a pure state can be
expressed in the following form:
\begin{equation*}
|\psi(t) \rangle = \sum_k c_k(t) |k\rangle,
\end{equation*}
where the coefficients $\{c_k\}$ satisfy the normalization condition
$\sum_k |c_k(t)|^2 \,{=}\, 1$. One can reduce this expression to the form of a projective vector by introducing the complex coordinates $z_k\,{=}\,c_k/c_1$ for $k\,{=}\,2$, \ldots, $n$ which parameterize the maximal
degenerate orbit where the state $|\psi\rangle$ lives, and $\hat{z}\,{\in}\,Z$ has the following form
\begin{equation*}
\hat{z} = \begin{pmatrix}
1 & 0 & 0 & \dots & 0 \\ z_2 & 1 & 0 &\dots & 0
\\ z_3 & 0 & 1 & \dots & 0  \\ \dots&\dots&\dots&\dots&\dots \\ z_n & 0 & 0 & \dots & 1
\end{pmatrix}.
\end{equation*}
Let $\hat{r}$ and $\hat{a}$ have the form
\begin{eqnarray*} \hat{r} =  \begin{pmatrix}
1 & r_{12} & r_{13} & \dots & r_{1n} \\ 0 & 1 & r_{23} &\dots & r_{2n}
\\ 0 & 0 & 1 & \dots & r_{3n}  \\ \dots&\dots&\dots&\dots&\dots \\ 0 & 0 & 0 & \dots & 1
\end{pmatrix},\\ \hat{a} =  \begin{pmatrix} a_2 & 0 & 0 & \dots & 0 \\
 0& a_3/a_2 & 0 & \dots & 0 \\ 0&0&a_4/a_3&\dots&0 \\ \dots&\dots&\dots&\dots&\dots \\
 0&0&0&\dots &1/a_{n} \end{pmatrix}
\end{eqnarray*}
After solving \eqref{IwasawaCalc} we obtain the expressions
\begin{eqnarray*}
a_k^2 = 1+\sum_{j=k}^n |z_j|^2,\quad r_{1k} = \bar{z}_k /a^2_{2},\ \ k\geqslant 2,
\\ r_{jk} = -z_j \bar{z}_k /a^2_{j+1},\ \ k\geqslant j+1,
\end{eqnarray*}
which provide a parameterization of the matrices
$\hat{a}$ and~$\hat{r}$.

The matrix $\hat{u}$ calculated by \eqref{IwasawaDec} reallizes evolution of the pure
state $|\psi\rangle$ starting from the initial state $|\psi(t_0)\rangle\,{=}\,|1\rangle$. In the case of other
initial state $|\psi_0 \rangle$, one should multiply $\hat{u}$ by the matrix $\hat{u}_0^{-1}$
such that $\hat{u}_0$ realizes the evolution form $|1\rangle$ to $|\psi_0 \rangle$.
Having the evolution operator $\hat{U}_{t,t_0}\,{=}\,\hat{u}\hat{u}_0^{-1}$,
one computes the corresponding Hamiltonian by the formula
\begin{equation*}
\hat{H}\,{=}\, i\hbar \frac{d \hat{U}_{t,t_0} }{dt} \hat{U}^{-1}_{t,t_0} =
i\hbar \frac{d \hat{u} }{dt} \hat{u}^{-1},
\end{equation*}
which does not depend on the
initial state.

The obtained evolution operator $\hat{U}_{t,t_0}$ realizes exactly the prescribed evolution of $|\psi\rangle$,
but is not a unique solution of the problem in question. According to Remark~\ref{r:1} any evolution operator
$\hat{U}_{t,t_0}\,{=}\,\hat{u}\hat{v}\hat{u}_0^{-1}$, where $\hat{v}\,{\in}\, \mathrm{U}(1)\,{\times}\,\mathrm{SU}(n\,{-}\,1)$, realizes the same evolution of $|\psi\rangle$. If $\hat{v}$ is a function in $t$, the corresponding Hamiltonian changes into
\begin{equation}\label{GaugeTransH}
\hat{H}_{\hat{v}}\,{=}\, \hat{H} +
i\hbar \hat{u} \frac{d \hat{v}}{dt}
\hat{v}^{-1} \hat{u}^{-1},
\end{equation}
which is a gauge transformation of $\hat{H}$.

\subsection{Mixed states}
Considering mixed states we represent them by density matrices, which are Hermitian positive definite matrices of unit trace. Because an orbit is specified by a diagonal matrix, one should diagonalize the given matrix~$\hat{\rho}$:
\begin{equation*}
\hat{\rho} = \hat{D} \hat{\rho}_{\text{in}} \hat{D}^{-1},
\end{equation*}
where $\hat{\rho}_{\text{in}}$ is the canonical form of $\hat{\rho}$.
As proven at the end of section~\ref{ss:OrbitStruct}, an orbit structure of the state space implies that the matrix $\hat{\rho}_{\text{in}}$ is constant and serves as the canonical form of $\hat{\rho}(t_0)$. Preserving the Hermitian character of $\hat{\rho}$, the matrix $\hat{D}$ should be a unitary operator, so it can serve as the desired evolution operator realizing the given evolution $\hat{\rho}$.
Here we show how to obtain such operator by means of the proposed method.

We start with constructing a generic element
the group $\mathrm{SU}(n)$ according to the scheme of
subsection~\ref{ss:StereogrProj} from the matrix
\begin{equation*}
\hat{z} = \begin{pmatrix}
1 & 0 & 0 & \dots & 0 \\ z_{12} & 1 & 0 &\dots & 0
\\ z_{13} & z_{23} & 1 & \dots & 0  \\ \dots&\dots&\dots&\dots&\dots \\
z_{1n} & z_{2n} & z_{3n} & \dots & 1 \end{pmatrix}.
\end{equation*}
In order to find  relations between the entries of $\hat{\rho}$ and the canonical coordinates
$\{z_{jk}\}$ we solve the equation
\begin{equation*}
\hat{\rho} = \hat{u} \hat{\rho}_{\text{in}}  \hat{u}^{-1}
\end{equation*}
for $\{z_{jk}\}$.
The relations allow to introduce the prescribed time evolution into the matrix $\hat{u}$.
The obtained time dependent matrix $\hat{u}$ realizes the evolution starting from $\hat{\rho}_{\text{in}}$, If $\hat{\rho}(t_0)$ does not coincide with its canonical form~$\hat{\rho}_{\text{in}}$,   the matrix $\hat{u}$ should be multiplied by the matrix $\hat{u}_0^{-1}$ such that $\hat{\rho}(t_0)\,{=}\,\hat{u}_0\hat{\rho}_{\text{in}}\hat{u}_0^{-1}$.
Finally, the given evolution from $\hat{\rho}(t_0)$ to $\hat{\rho}(t)$ is realized by the operator $\hat{U}_{t,t_0} \,{=}\, \hat{u} \hat{u}_0^{-1}$. Evidently, the Hamiltonian does not depend on $\hat{u}_0$:
\begin{equation*}
\hat{H}\,{=}\, i\hbar \frac{d \hat{U}_{t,t_0}}{dt} \hat{U}^{-1}_{t,t_0} =
i\hbar \frac{d \hat{u} }{dt} \hat{u}^{-1}.
\end{equation*}

We obtain the Hamiltonian model guaranteeing that $\hat{\rho}$ evolves according to the prescribed law. Again this model is not unique. Any evolution operator $\hat{U}_{t,t_0} \,{=}\, \hat{u}  \hat{v} \hat{u}_0^{-1}$ with the matrix $\hat{v}$ belonging to the stability subgroup of the orbit where the state $\hat{\rho}$ lives also realizes the prescribed evolution of $\hat{\rho}$. In this case $\hat{v}$ performs a gauge transformation of the model, where the Hamiltonian changes according to the law \eqref{GaugeTransH}.

\section{Illustration of the method}
To illustrate the method proposed in the previous section we
consider two examples: a pure state and a mixed one. In the both cases
we calculate evolution operators and the corresponding Hamiltonians,
and show how to construct other possible evolution operators.

\subsection{The case of a pure state}
As a time dependent normalized pure state we choose the following one\cite{Leggio}
\begin{eqnarray}\label{StateExmpl}
|\psi(t)\rangle = \frac{1}{\sqrt{2+\cos^2\omega t}}
\Bigl\{|1,-1\rangle +\cos\omega t
|0,0\rangle + \nonumber \\  e^{i\pi/6}|{-}1,1\rangle\Bigr\},
\end{eqnarray}
describing a time evolution of a pair
of interacting spins~1: $\bm{S}_1$ and
$\bm{S}_2$. The state $|m_1,m_2\rangle$ is a
simultaneous eigenstate of $S_{1z}$ and
$S_{2z}$, that is
$S_{iz}|m_1,m_2\rangle\,{=}\,m_i|m_1,m_2\rangle$, where
$m_i\,{=}\,{-}1,\,0,\,1$ for $i\,{=}\,1,\,2$.

It is worth noting that the evolution \eqref{StateExmpl} of $|\psi(0)\rangle$
exhibits the conservation of $m\,{=}\,(m_1\,{+}\,m_2)\,{=}\,0$, which is an eigenvalue
of the observable $S_z\,{=}\,S_{1z}\,{+}\,S_{2z}$.
As a result, $|\psi\rangle$, which is indeed a vector from the
nine-dimensional Hilbert space $\mathcal{H}_{12}$ of the two
qutrits, lives in a three-dimensional subspace
$\mathcal{H}_0\,{\subset}\,\mathcal{H}_{12}$ of
$S_z$ pertaining to $m\,{=}\,0$. Thus, we need not to take into consideration the whole
Hilbert space $\mathcal{H}_{12}$. It is enough to describe the evolution of $|\psi\rangle$
by a reduced Hamiltonian model living in $\mathcal{H}_0$.

Another interesting observation is the following.
Focusing on the mathematical nature of the problem we could replace
the specific evolution \eqref{StateExmpl} of two qutrits by the evolution of a single qutrit
$\bm{J}$ such that $J_z|M\rangle\,{=}\,M|M\rangle$ with
$M\,{=}\,0,\,{\pm}1$:
\begin{equation}\label{StateExmpl_red}
|\psi(t)\rangle = \frac{1}{\sqrt{2+\cos^2\omega t}} \left\{|1\rangle
+\cos\omega t |0\rangle + e^{i\pi/6}|{-}1\rangle\right\}.
\end{equation}
Both evolutions: \eqref{StateExmpl} and \eqref{StateExmpl_red}
can be described by the same Hamiltonian model constructed below.

For the sake of generality we consider a state of the form
\begin{equation}\label{State}
|\psi(t)\rangle = c_1(t)|1,{-}1\rangle + c_{2}(t) |0,0\rangle +
c_3(t) |{-}1,1\rangle
\end{equation}
with $|c_1(t)|^2+|c_2(t)|^2+|c_3(t)|^2=1$. While \eqref{State} is a
pure state, its evolution is located within a maximum degenerate orbit
$\mathrm{SU}(3)/(\mathrm{U}(1)\,{\times}\, \mathrm{SU}(2))$, which
is the projective space $\CP^2$. Reduction to the form of a
projective vector gives
\begin{equation*}
|\psi(t)\rangle = |1,{-}1\rangle + z_{2}(t) |0,0\rangle + z_3(t)
|{-}1,1\rangle,
\end{equation*}
where $z_2\,{=}\,c_2/c_1$, $z_3\,{=}\,c_3/c_1$. On the other hand,
this projective vector can be rearranged into
the matrix form:
\begin{equation*}
\hat{z} = \begin{pmatrix} 1&0&0\\
z_2&1&0 \\ z_3&0&1 \end{pmatrix}
\end{equation*}

From \eqref{IwasawaCalc} we obtain the following:
\begin{eqnarray*}
&a_2^2 = 1+|z_2|^2+|z_3|^2,\quad r_{12} = \frac{\bar{z}_2}{a_2^2},\quad r_{13} = \frac{\bar{z}_3}{a_2^2},\\
&a_3^2 = 1+ |z_3|^2,\quad r_{23} = -\frac{z_2 \bar{z}_3}{a_3^2}.
\end{eqnarray*}
Then an evolution matrix $\hat{u}$ looks like
\begin{equation}\label{UMatr}
\hat{u} = \hat{z} \hat{r}^{-1} \hat{a}^{-1} =
\begin{pmatrix} \frac{1}{a_2}&\frac{-\bar{z}_2}{a_2 a_3}& \frac{-\bar{z}_3}{a_3}\\
\frac{z_2}{a_2}&\frac{a_3}{a_2}&0 \\ \frac{z_3}{a_2}& \frac{-z_3
\bar{z}_2}{a_2 a_3}&\frac{1}{a_3} \end{pmatrix}.
\end{equation}
This matrix acting on the vector $\bm{e} \,{=}\,
(1,\,0,\,0)$ gives $\bm{z} \,{=}\,
\bigl(1/a_2,\,z_2/a_2,\,z_3/a_2\bigr)$, where $a_2$ serves as its
norm. For an arbitrary initial vector
$\bm{z}_0$ we need to find the matrix $\hat{u}_0$ such that $\hat{u}_0\bm{e}
\,{=}\, \bm{z}_0$. Then
\begin{equation*}
\bm{z} = \hat{u} \hat{u}_0^{-1} \bm{z}_0,
\end{equation*}
and $\hat{U}_t\,{=}\,\hat{u} \hat{u}_0^{-1}$ is the required evolution operator.

Now we come back to the state \eqref{StateExmpl}. In this case,
$z_2(t) \,{=}\, \cos \omega t$, $z_3(t) \,{=}\, e^{i\pi/6}$, and
$a_2^2 = 2+\cos^2 \omega t$, $a_3^2 \,{=}\, 2$. Then
\begin{eqnarray*}
&\hat{u} = \begin{pmatrix}\frac{1}{\sqrt{2+\cos^2 \omega t}} &
\frac{-\cos \omega t}{\sqrt{2(2+\cos^2 \omega t)}} & \frac{-e^{-i\pi/6}}{\sqrt{2}}
\\ \frac{\cos \omega t}{\sqrt{2+\cos^2 \omega t}} & \frac{\sqrt{2}}{\sqrt{2+\cos^2 \omega t}} &
0 \\ \frac{e^{i\pi/6}}{\sqrt{2+\cos^2 \omega t}} & \frac{-e^{i\pi/6}\cos\omega t}{\sqrt{2(2+\cos^2 \omega t)}}
& \frac{1}{\sqrt{2}} \end{pmatrix},\\
&\hat{u}_0 = \begin{pmatrix} \frac{1}{\sqrt{3}} &
\frac{-1}{\sqrt{6}} & \frac{-e^{-i\pi/6}}{\sqrt{2}}
\\ \frac{1}{\sqrt{3}} & \frac{\sqrt{2}}{\sqrt{3}} &
0 \\ \frac{e^{i\pi/6}}{\sqrt{3}} & \frac{-e^{i\pi/6}}{\sqrt{6}}
& \frac{1}{\sqrt{2}} \end{pmatrix},
\end{eqnarray*}
and the evolution matrix is
\begin{widetext}
\begin{equation*}
\small \hat{U}_t = \begin{pmatrix}
\frac{2+\cos \omega t + \sqrt{3(2+\cos^2 \omega t)}}{2\sqrt{3(2+\cos^2 \omega t)}} & \frac{1-\cos \omega t}{\sqrt{3(2+\cos^2 \omega t)}} &
\frac{2+\cos\omega t - \sqrt{3(2+\cos^2 \omega t)}}{2\sqrt{3(2+\cos^2 \omega t)}}\, e^{-i\pi/6} \\
\frac{\cos \omega t - 1}{2\sqrt{3(2+\cos^2 \omega t)}} & \frac{2+\cos \omega t}{\sqrt{3(2+\cos^2 \omega t)}} & \frac{\cos \omega t - 1}{2\sqrt{3(2+\cos^2 \omega t)}}\, e^{-i\pi/6} \\
\frac{2+\cos \omega t - \sqrt{3(2+\cos^2 \omega
t)}}{2\sqrt{3(2+\cos^2 \omega t)}}\,e^{i\pi/6} &
\frac{1-\cos \omega t} {\sqrt{3(2+\cos^2 \omega
t)}}\,e^{i\pi/6} & \frac{2+\cos \omega t + \sqrt{3(2+\cos^2 \omega
t)}}{2\sqrt{3(2+\cos^2 \omega t)}} \end{pmatrix}.
\end{equation*}
\end{widetext}
The Hamiltonian calculated by \eqref{HamDef} is the following
\begin{equation}\label{H}
\hat{H} = \frac{i \hbar \omega \sin \omega t}{2+\cos^2 \omega t}
\begin{pmatrix} 0& 1 & 0 \\ -1 & 0 & -e^{i \pi /6} \\
0&e^{-i\pi/6} & 0\end{pmatrix},
\end{equation}
and coincides with the result reported in \cite{Leggio}.

As mentioned in Remark~\ref{r:1}, the obtained evolution matrix $\hat{U}_t$ is not a unique solution of the problem. As well the problem is solved by any matrix $\hat{U}_t\,{=}\,\hat{u} \hat{v} \hat{u}_0^{-1}$ with $\hat{v}$ of the following form
\begin{equation*}
\hat{v} =
\begin{pmatrix} e^{i\gamma(t)/\hbar} & 0 & 0 \\ 0 & \alpha(t) & \beta(t) \\
0& -\beta^{\ast}(t) & \alpha^{\ast}(t)\end{pmatrix},
\end{equation*}
where $\gamma$ is an arbitrary real-valued function in $t$, and $\alpha$, $\beta$ are complex-valued functions in $t$ such that $|\alpha(t)|^2\,{+}\,|\beta(t)|^2\,{=}\,1$ for all $t$.

It is easy to check that the reduced to $\mathcal{H}_0$ operators $\hat{S}_{1z}\,{+}\,\hat{S}_{2z}$, $\hat{S}_{1z}\hat{S}_{2z}$,
$\hat{S}_1^+\hat{S}_2^-$, and $\hat{S}_1^-\hat{S}_2^+$, where $\hat{S}_i^{\pm}\,{=}\,\hat{S}_{ix}\pm i \hat{S}_{iy}$, $i\,{=}\,1,\,2$,
can be cast in the following matrix form:
\begin{eqnarray}
&\hat{S}_1^z+\hat{S}_2^z=
\begin{pmatrix}
0 & 0 & 0\\
0 & 0 & 0\\
0 & 0 & 0\\
\end{pmatrix},\quad
\hat{S}_1^z \hat{S}_2^z=
\begin{pmatrix}
-1 & 0 & 0\\
0 & 0 & 0\\
0 & 0 & -1\\
\end{pmatrix},  \nonumber \\
&\hat{S}_1^+ \hat{S}_2^-=
\begin{pmatrix}
0 & 2 & 0\\
0 & 0 & 2\\
0 & 0 & 0\\
\end{pmatrix},\quad
\hat{S}_1^-\hat{S}_2^+=
\begin{pmatrix}
0 & 0 & 0\\
2 & 0 & 0\\
0 & 2 & 0\\
\end{pmatrix}. \label{SS}
\end{eqnarray}
Exploiting the expressions \eqref{SS} a Hamiltonian model $\hat{H}_{\rm gen}$ acting in $\mathcal{H}_{12}$,
whose reduction to $\mathcal{H}_0$ is given by \eqref{H}, is written in the
following form
\begin{eqnarray}\label{H1}
&\hat{H}_{\rm gen}=\hbar\Omega_0(\hat{S}_1^z+\hat{S}_2^z)+ \nonumber \\ &i \hbar \lambda(t)\Bigl( (e^{-i\frac{\pi}{6}}\hat{S}_1^+\hat{S}_2^-+\hat{S}_1^-\hat{S}_2^+)\hat{S}_1^z \hat{S}_2^z - \nonumber\\
&\hat{S}_1^z \hat{S}_2^z(\hat{S}_1^+\hat{S}_2^-+e^{i\frac{\pi}{6}}\hat{S}_1^-\hat{S}_2^+)\Bigr)
\end{eqnarray}
where the coupling constant $\lambda$ has the form
\begin{equation}
\lambda(t)=-\frac{1}{2}\cdot \frac{\omega\sin\omega t}{2+\cos^2\omega t}
\end{equation}
and $\hbar\Omega_0$ is an arbitrary value of energy. Equation \eqref{H1} may
be interpreted as describing two interacting
qutrits $\bm{S}_1$ and $\bm{S}_2$ subjected to a
magnetic field. The Hamiltonian $\hat{H}_{\rm gen}$ commutes with $\hat{S}_z$ at any
$t$ and guarantees that the value of $m\,{=}\,(m_1\,{+}\,m_2)$ remains constant, thus the evolution of $|\psi(0)\rangle$
belongs to a maximal degenerate orbit of SU(3) over the state space.

\subsection{The case of mixed state}\label{Ex2}
To illustrate our method in the case of a mixed
state, let us consider a two-level atom coupled to a bosonic mode.
Atomic dynamical variables can be represented in terms of the
Pauli matrices $\sigma_x,\,\sigma_y,\,\sigma_z$ acting upon a
two-dimensional Hilbert space where a basis is
$\{|+\rangle,|-\rangle\}$ with $\sigma_z|\pm\rangle=\pm|\pm\rangle$.
The bosonic mode is represented, as usual, by an annihilation $a$ and
a creation $a^{\dagger}$ operators
such that $[a,a^{\dagger}]=1$, acting upon an
infinite-dimensional Hilbert space with the basis  $\{|n\rangle,
n=0,1,2,\ldots\}$ such that $a^{\dagger}a|n\rangle=n|n\rangle$. Suppose
that at $t=0$ the composed system is in the mixed state
\begin{equation}
\rho(0)=(3/4)|0,+\rangle\langle0,+| + (1/4)|1,-\rangle\langle1,-|,
\end{equation}
where $|n,\pm\rangle$ is a simultaneous eigenstate of
$a^{\dagger}a$ and $\hat{\sigma}_z$ pertaining to their eigenvalues $n$
and $\pm1$ respectively. Assume that the atom-mode interaction
 generates the following evolution:
\begin{eqnarray}\label{rho}
&\rho(t)=\Bigl(3/4-(1/2)\cos^2 gt \Bigr) |0,+\rangle\langle0,+|
 +  \nonumber \\ &\Bigl(1/4+(1/2)\cos^2 gt \Bigr)|1,-\rangle\langle1,-| +  \nonumber \\
&(1/4)\sin 2gt \Bigl(|0,+\rangle\langle1,-|+ |1,-\rangle\langle0,+|\Bigr).
\end{eqnarray}

It is suitable to represent $\rho$ in a matrix form as follow:
\begin{equation}\label{MStateExmpl}
\hat{\rho}(t)=\begin{pmatrix} 3/4 - (1/2) \cos^2 gt & (1/4) \sin 2 gt  \\
(1/4) \sin 2 gt  & 1/4 + (1/2) \cos^2 gt \end{pmatrix},
\end{equation}
which may be generalized by the following notation for entries of
$\hat{\rho}$:
\begin{equation*}
\hat{\rho}=\begin{pmatrix} \rho_{11}&\rho_{12} \\
\bar{\rho}_{12} & \rho_{22} \end{pmatrix},
\end{equation*}
Diagonalization of $\hat{\rho}$ leads to
\begin{eqnarray*}
&\hat{\rho}_{\text{in}}=\begin{pmatrix} \lambda_{1}&0 \\
0 & \lambda_{2} \end{pmatrix},\\ &\lambda_{1,2} = \frac{1}{2}\left(\rho_{11}+\rho_{22} \pm
\sqrt{(\rho_{11}-\rho_{22})^2+4|\rho_{12}|^2}\right).
\end{eqnarray*}
The group acting over the state \eqref{MStateExmpl} is $\mathrm{SU}(2)$,
and its generic element obtained by the explained above method of stereographic parametrization has the form
\begin{equation*}
\hat{u}=\frac{1}{\sqrt{1+|z|^2}} \begin{pmatrix} 1&-\bar{z} \\
z & 1 \end{pmatrix}.
\end{equation*}
Solving the equation $\hat{u}\hat{\rho}_{\text{in}}\hat{u}^{-1} \,{=}\, \hat{\rho}$
for $z$, we come to
\begin{equation*}
z = \frac{2\bar{\rho}_{12}}{\rho_{11}-\rho_{22} \pm
\sqrt{(\rho_{11}-\rho_{22})^2+4|\rho_{12}|^2}}.
\end{equation*}

In the case of state \eqref{MStateExmpl} we obtain
\begin{equation*}
\hat{\rho}_{\text{in}} = \begin{pmatrix} 1/4&0 \\
0 & 3/4 \end{pmatrix},
\end{equation*}
and $z \,{=}\, -\tan gt$. Then a possible evolution  operator
and its correspondent Hamiltonian acquire the forms
\begin{eqnarray}\label{U}
\hat{U}_t = \begin{pmatrix} \cos gt &  \sin gt \\
-\sin gt & \cos gt \end{pmatrix},\quad
\hat{H} =   i \hbar g \begin{pmatrix} 0 & 1 \\
-1 &0\end{pmatrix}.
\end{eqnarray}
The given evolution \eqref{MStateExmpl} is obtained by
\begin{equation*}
\hat{\rho}(t) = \hat{U}_t \hat{\rho}(0) \hat{U}_{t}^{-1}.
\end{equation*}

According to Remark~\ref{r:1} there exist other solutions of the problem.
Evolution operators $\hat{U}_t \,{=}\, \hat{u}\hat{v}$ such that
\begin{equation*}
\hat{v}=\begin{pmatrix} e^{i\alpha(t)/\hbar} & 0 \\
0 & e^{i \beta(t)/\hbar}\end{pmatrix},
\end{equation*}
where $\alpha$ and $\beta$ are arbitrary real-valued functions in $t$, also realizes
the same evolution~\eqref{MStateExmpl}.
The corresponding Hamiltonian $\hat{H}$ gets the form
\begin{eqnarray*}
\hat{H} = &  i \hbar g \begin{pmatrix} 0 & 1 \\
-1 &0\end{pmatrix} -  \frac{1}{2} \Bigl(\alpha'(t) + \beta'(t) \Bigr)
\begin{pmatrix} 1&0\\0&1 \end{pmatrix} + \\ & \frac{1}{2} \Bigl(\alpha'(t) - \beta'(t) \Bigr)
\begin{pmatrix} - \cos 2gt
 & \sin 2gt \\ \sin 2gt & \cos 2gt \end{pmatrix}.
\end{eqnarray*}

Once again we emphasize that although the Hilbert space of the
composite system is infinite-dimensional, the prescribed time
evolution belongs to the state space
$\mathcal{P}$ over the two-dimensional invariant subspace
$\Hilbert_1$ of the excitation number
operator $E\,{=}\,a^{\dagger}a\,{+}\,(1/2)\sigma_z + 1/2$, where only one
excitation is present with certainty. In this
physical context our method generates a Hamiltonian model reduced to
$\mathcal{P}$. It is easy to check that a
reduction of the time-independent Hamiltonian model
\begin{equation}\label{JC}
H_{\rm JC} = \hbar g(a^{\dagger}\sigma_-+a\sigma_+)
\end{equation}
has a matrix representation coincident with the model given by
(\ref{U}). The Hamiltonian $H_{\rm JC}$
represents the well known Jaynes-Cummings model of a two-level atom
resonantly coupled to a single electromagnetic field mode, taken in
the interaction picture with respect to
$H_0\,{=}\,\hbar\omega\left(a^{\dagger}a+\frac{1}{2}\sigma_z\right)$. Even
in this example, the same mathematical construction would be valid
if we substitute $|0,+\rangle\langle0,+|$ and
$|1,-\rangle\langle1,-|$ in (\ref{rho}) with
the two basis projectors $|+\rangle\langle+|$ and
$|-\rangle\langle-|$ of a single qubit. In
this case (\ref{U}) would describe the interaction of the qubit with
a particular magnetic field.

\section{Conclusion and discussion}
In the paper we have proposed  a method of reconstructing a Hamiltonian from a prescribed time evolution of a quantum state $\rho$. We~suppose that the quantum system is governed by the von Neumann equation, and so every evolution is represented by a unitary operator. This implies that the action of the group of evolution operators over the state space is equivalent to the adjoint action of the unitary group over Hermitian matrices. As a result the conditions $\Tr \rho^k(t)\,{=}\,{\rm const}$ for all  $k\,{\geqslant}\, 1$, determining an orbit of the adjoint action of the unitary group, are fulfilled. For every such state $\rho$ it is possible to construct
the class of evolution operators realizing the prescribed evolution, and this evolution is located within a single orbit.

In  the theory of open quantum systems
one deals with reduced density matrices. They are governed by more
complicate equations than the von Neumann one. These equations,
known as quantum master equations, are different for different
systems. Evidently, the dynamics of an open quantum
system, in general, can not be represented by a unitary operator.
Nevertheless, we hope that the inverse dynamical problem for reduced density matrices may be  solved within the
group-theoretical approach reported in this paper.

\begin{acknowledgments}
We would like to thank professor P.~Holod for fruitful discussion and pieces of sensible advice during preparation of the paper.
One of us (AM) thanks Elena Ferraro and Bruno Leggio for help
in the preparation of this manuscript and
scientific discussions.

This work is supported by the grant
MIUR-Cooperazione interuniversitaria internazionale --- capitolo 1706/6 --- A.F. 2007,  and a grant of the International Charitable Fund for Renaissance of Kiev-Mohyla Academy.
\end{acknowledgments}

\nocite{*}
\providecommand{\noopsort}[1]{}\providecommand{\singleletter}[1]{#1}%

\end{document}